# Erratum to Three Papers on the *β*-Pyrochlore Oxide Superconductors

Zenji Hiroi, Shigeki Yonezawa, Jun-Ichi Yamaura, Takaki Muramatsu,
Yoshitaka Matsushita and Yuji Muraoka

*Institute for Solid State Physics, University of Tokyo, Kashiwa, Chiba 277-8581*

The magnitude of magnetic fields given in the three papers[1-3] that report on the superconductivity of the *β*-pyrochlore oxides is incorrect due to a fatal error in our experimental equipment (Physical Property Measurement System (PPMS), Quantum Design). Recently, we noticed that the magnet controller of the PPMS had been supplying only 58% of the expected current to the superconducting magnet. Thus, all the field values given in the previous papers[1-3] must be reduced by a factor of 0.58: the maximum field of 14 T is actually about 8 T, for instance. After the recovery of the magnet controller we measured the field dependence of the superconducting transition temperature $T_c$ on a single crystal of $KOs_2O_6$, and found it to be in good agreement with the previous data when it is reduced by this factor. Figure 1 shows a revised magnetic field versus temperature phase diagram for $KOs_2O_6$, which should replace Fig. 4 of reference 3. Note that the slopes of the two upper critical field $H_{c2}$ curves are nearly equal between the new data (red cross) and the previous data after the correction (blue circle).

On $RbOs_2O_6$ the $H_{c2}$ value estimated at $T = 0$ K[1] should be reduced to about 10 T; this was determined from the field dependence of $T_c$ defined as the midpoint of a resistive transition. On the other hand, a much smaller value of 6 T was reported by Brühwiler *et al.* on the basis of their specific heat measurements.[4] Thus, there still remains inconsistency in the $H_{c2}$ value of $RbOs_2O_6$. Reexamination of this issue will be carried out and reported elsewhere. Aside from the magnitude of the magnetic fields, nothing in discussions nor the conclusions given in the three papers is altered by this error.

3) Z. Hiroi, S. Yonezawa, J. Yamaura, T. Muramatsu and Y. Muraoka: J. Phys. Soc. Jpn. **74** (2005) 1682.

4) M. Brühwiler, S. M. Kazakov, N. D. Zhigadlo, J. Karpinski and B. Batlogg: Phys. Rev. B **70** (2004) 020503R.

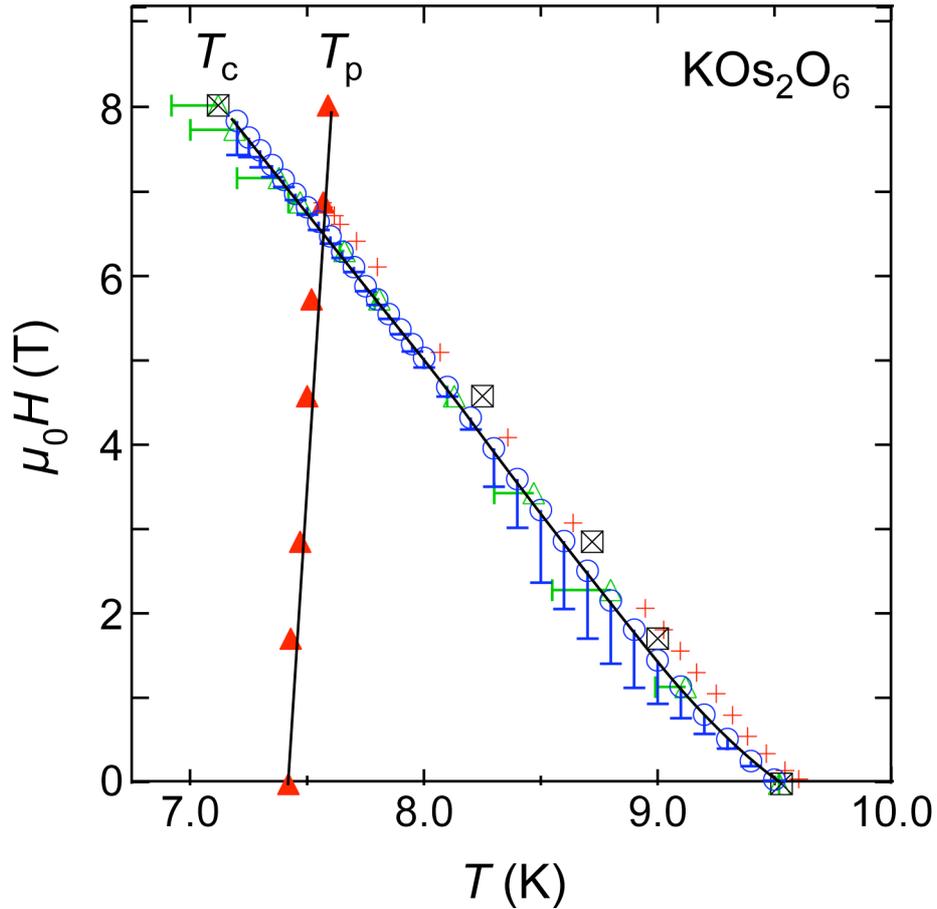

Fig. 1. (Color online) Revised magnetic-field versus temperature-phase diagram for $KOs_2O_6$, which should replace Fig. 4 of reference 3. Open circles, triangles, and squares are the $H_{c2}$ values determined by resistivity, specific heat, and magnetization measurements, respectively.[3] Red crosses represent new data obtained by resistivity measurements on another single crystal. Bars from the data marks represent the extension of the superconducting transition. Solid triangles indicate the second peak temperatures. Solid lines serve as visual guides.